\documentclass{appolb}

\def\R{{\rm R}}
\def\Li{{\rm Li}}
\def\z#1{{\zeta(#1)}}
\def\RR#1{{\rm R_{#1}}}
\def\l{\langle}
\def\r{\rangle} 
\def\eps{\varepsilon}

\def\kp{p_4}
\def\km{p_5}
\def\nf{n_f}

\def\qb{\bar q}

\def\x#1{{x_{#1}}}
\def\nn{\nonumber}

\def\as{{\alpha_s}}

\def\ren{{\rm ren}}
\def\fin{{\rm fin}}

\begin{document}
\pagestyle{plain}


\newcount\eLiNe\eLiNe=\inputlineno\advance\eLiNe by -1
\title{TWO-LOOP AMPLITUDES FOR $ e^+ e^- \to q \qb g$: \\
  THE $n_f$-CONTRIBUTION}

\author{Sven Moch$^a$, Peter Uwer$^a$ and Stefan Weinzierl$^b$
\address{
$^a$Institut f{\"u}r Theoretische Teilchenphysik, 
  Universit{\"a}t Karlsruhe \\ 76128 Karlsruhe, Germany \\[1mm]
$^b$Dipartimento di Fisica, Universit\`a di Parma \\
       INFN Gruppo Collegato di Parma, 43100 Parma, Italy \\
}}

\maketitle

\begin{abstract}
    We discuss the calculation of the 
    $n_f$-contributions to the two-loop 
    amplitude for $ e^+ e^- \rightarrow q g \bar{q}$. 
    The calculation uses an efficient method based on nested sums. 
    The result is presented in terms of multiple polylogarithms 
    with simple arguments, which allow for analytic continuation in a 
    straightforward manner.
\end{abstract}

\section{Introduction}
Searches for new physics in particle physics rely to a large extend 
on our ability to constrain the parameters of the standard model.
For instance, the strong coupling constant $\alpha_s$ can be measured 
by using the data for $e^+ e^- \rightarrow \mbox{3-jets}$.
At present, the error on the extraction of $\alpha_s$ from this measurement
is dominated by theoretical uncertainties \cite{Dissertori:2001mv}, 
most prominently, by the truncation of the perturbative expansion at a 
fixed order.

The perturbative QCD calculation of $e^+ e^- \rightarrow \mbox{3-jets}$ at
next-to-next-to-leading order (NNLO) 
requires the tree-level amplitudes for 
$e^+ e^- \rightarrow \mbox{5 partons}$ \cite{Berends:1989yn}, 
the one-loop amplitudes for
$e^+ e^- \rightarrow \mbox{4 partons}$ \cite{Bern:1997ka,Glover:1997eh}
as well as the two-loop amplitude 
for $e^+e^-\to q \qb g$ together with the one-loop
amplitude $e^+e^-\to q \qb g$ to order $\eps^2$ 
in the parameter of dimensional regularization.

The helicity averaged squared matrix elements at the
two-loop level for $e^+e^-\to q \qb g$ have recently been given \cite{Garland:2001tf}.
In contrast, having the two-loop amplitude available, one 
keeps the full correlation between the incoming $e^+e^-$
and the outgoing parton's spins and momenta. Thus, 
one can study oriented event-shape observables. 
In addition, one has the option to investigate event-shape observables 
in polarized $e^+e^-$-annihillation at a future linear $e^+e^-$-collider TESLA.

\section{Calculation}
\label{sec:calculation}

We are interested in the following reaction
\begin{equation}
  e^+ + e^- \rightarrow  q + g + \bar{q}\, ,
\end{equation}
which we consider in the form, 
$
  0   \rightarrow  q(p_1) + g(p_2) + \bar{q}(p_3) + e^-(\kp) +  e^+(\km) , 
$
with all particles in the final state, 
to be consistent with earlier work \cite{Bern:1997ka}. 
The kinematical invariants for this reaction are denoted by
\begin{eqnarray}
  s_{ij} = \left( p_i + p_j \right)^2, \;\;\; 
  s_{ijk} = \left( p_i + p_j + p_k \right)^2,\;\;\;
  s=s_{123}\, ,
\end{eqnarray}
and it is convenient to introduce the dimensionless quantities
\begin{eqnarray}
\label{eq:xidef}
  x_1 = \frac{s_{12}}{s_{123}}, \;\;\; x_2 = \frac{s_{23}}{s_{123}}.
\end{eqnarray}

Working in a helicity basis, it suffices to consider 
the pure photon exchange amplitude $\cal A_\gamma$ as it allows the reconstruction 
of the full amplitude with $Z$-boson exchange by adjusting the couplings.
Furthermore, the complete information about $\cal A_\gamma$ is given 
by just one independent helicity amplitude, which we take to be 
$A_\gamma(1^+, 2^+, 3^-, 4^+, 5^-)$. All other helicity configurations can be 
obtained from parity and charge conjugation. 

We can write $A_\gamma(1^+, 2^+, 3^-, 4^+, 5^-)$ in terms of 
coefficients $c_2,c_4,c_6$ and $c_{12}$ for the various independent spinor
structure as 
\begin{eqnarray}
\lefteqn{
  A_\gamma(1^+, 2^+, 3^-, 4^+, 5^-) \, = \, \frac{i}{\sqrt{2}} \frac{ [12]}{s^3} } 
\\ & & 
  \times \left\{
    s \l 35 \r [42] \left[ (1-x_1) \left( c_2 + \frac{2}{x_2} c_6 - c_{12}
  \right) + (1-x_2) \left( c_4 - c_{12} \right) + 2  c_{12}
 \right]
  \right. \nn \\
  & & \left.
    - \l 31 \r [ 12 ] \left[ 
      [ 43 ] \l 35 \r \left( c_2 + \frac{2}{x_2} c_6 - c_{12} \right)
      + [ 41 ] \l 15 \r \left( c_4 - c_{12} \right)
    \right]
  \right\}\, , 
\nn
\end{eqnarray}
where we have introduced the short-hand notation for spinors of 
definite helicity,
$
|i \pm \rangle = | p_i \pm \rangle = u_\pm(p_i) = v_\mp(p_i)
$, 
$
\langle i \pm | = \langle p_i \pm | = \bar{u}_\pm(p_i) = \bar{v}_\mp(p_i),
$
and for the spinor products $\langle p q \rangle =  \langle p - | q + \rangle$ 
and $\left[ p q \right] =  \langle p + | q - \rangle$.

The coefficients $c_i$ depend on the $\x1$ and $\x2$ of eq.(\ref{eq:xidef}) 
and can be calculated in conventional dimensional regularization. 
To that end, we proceed as follows \cite{Moch:2001zr}-\cite{Moch:2002hm}.
In a first step, with the help of Schwinger parameters 
\cite{Tarasov:1996br},
we map the tensor integrals to combinations of scalar integrals in various
dimensions and with various powers $\nu_i$ of the propagators. 
For every basic topology, these scalar integrals can be written as nested sums 
involving $\Gamma$-functions.
The evaluation of the nested sums proceeds systematically with the
help of the algorithms of \cite{Moch:2001zr}, which rely on the
algebraic properties of the so called $Z$-sums,
\begin{eqnarray}
\label{defZsums}
  Z(n;m_1,...,m_k;x_1,...,x_k) & = & \sum\limits_{n\ge i_1>i_2>\ldots>i_k>0}
     \frac{x_1^{i_1}}{{i_1}^{m_1}}\ldots \frac{x_k^{i_k}}{{i_k}^{m_k}}\, .
\end{eqnarray}
By means of recursion the algorithms allow to solve the nested sums in terms 
of a given basis in $Z$-sums to any order in $\eps$.
$Z$-sums can be viewed as generalizations of harmonic sums
\cite{Vermaseren:1998uu}
and an important subset of $Z$-sums are 
multiple polylogarithms \cite{Goncharov}, 
\begin{eqnarray}
\label{multipolylog}
\mbox{Li}_{m_k,...,m_1}(x_k,...,x_1) & = & Z(\infty;m_1,...,m_k;x_1,...,x_k).
\end{eqnarray}
All algorithms for this procedure have been implemented in FORM~\cite{Vermaseren:2000nd} 
and in the GiNaC framework \cite{Bauer:2000cp,Weinzierl:2002hv}.
In this way, we could calculate all loop integrals contributing to the one-
and two-loop virtual amplitudes very efficiently in terms of multiple polylogarithms.

The perturbative expansion in $\alpha_s$ of the functions $c_i$ is defined through
\begin{eqnarray}
  c_i &=& \sqrt{4 \pi \alpha_s} \left(
    c_i^{(0)} 
    +\left( \frac{\alpha_s}{2\pi}\right) c_i^{(1)} 
    +\left( \frac{\alpha_s}{2\pi}\right)^2 c_i^{(2)}
    + O(\as^3) 
  \right) \, .
\end{eqnarray}
Then, after ultraviolet renormalization, the infrared pole structure of the 
renormalized coefficients $c_{i}^{\ren}$ agrees with the prediction made by 
Catani \cite{Catani:1998bh} using an infrared factorization formula.
We use this formula to organize the finite part into terms arising from
the expansion of the pole coefficients and a finite remainder,
\begin{eqnarray}
   c_{i}^{(2),\fin} &=& c_{i}^{(1),\ren} - {\bf I}^{(1)}(\eps) c_{i}^{(1),\ren}
   - {\bf I}^{(2)}(\eps) c_{i}^{(0)} \, , 
  \label{eq:def-results2}
\end{eqnarray}
for $i=\{2,4,6,12\}$, and with the one- and two-loop insertion operators 
${\bf I}^{(1)}(\eps)$ and ${\bf I}^{(2)}(\eps)$ given in \cite{Catani:1998bh}.

As an example, we present our result for $\nf N$-contribution to the finite part 
$c_{12}^{(2),\fin}$ at two loops,
\begin{eqnarray}
  &&c_{12}^{(2),\fin}(\x1,\x2) =
  \nf N \bigg(
  3 {\ln(\x1)\over (\x1\!+\!\x2)^2}
  +{1 \over 4}  {\ln(\x2)^2 - 2   \Li_{2}(1\!-\!\x2) 
 \over \x1 (1\!-\!\x2)}
  \\&&
  +{1 \over 12}  {\z2\over (1\!-\!\x2) \x1}
  -{1 \over 18}  {13 \x1^2\!+\!36 \x1\!-\!10 \x1 \x2\!-\!18 \x2\!+\!31 \x2^2
  \over (\x1\!+\!\x2)^2 \x1 (1\!-\!\x2)}
   \ln(\x2)
  \nn\\&&
  +{\x1^2\!-\!\x2^2\!-\!2 \x1\!+\!4 \x2 \over (\x1\!+\!\x2)^4} \RR1(\x1,\x2)
  -{1 \over 12}  {\R(\x1,\x2) \over \x1 (\x1\!+\!\x2)^2} \bigg[
  5 \x2+42 \x1+5
  \nn\\&&
  -{(1\!+\!\x1)^2\over 1\!-\!\x2}
  -4 {1\!-\!3 \x1\!+\!3 \x1^2\over 1\!-\!\x1\!-\!\x2}-72 {\x1^2\over \x1\!+\!\x2}\bigg]
  +\bigg[
   {1 \over 12}  {1\over \x1 (1\!-\!\x2)}
  +{6\over (\x1\!+\!\x2)^3}
  \nn\\&&
  -{1\!+\!2 \x1\over \x1 (\x1\!+\!\x2)^2}
  \bigg] (\Li_{2}(1\!-\!\x2)-\Li_{2}(1\!-\!\x1))
  -{1\over (\x1\!+\!\x2) \x1}
  \bigg)
  \nn\\&&
 -{1 \over 2}  I \pi \nf N {\ln(\x2) \over \x1 (1\!-\!\x2)}
  \nn \, .
\end{eqnarray}
We have introduced the function $\R(\x1,\x2)$, 
which is well known from \cite{Ellis:1981wv}, 
\begin{eqnarray}
  \label{eq:Rfunction}
\lefteqn{
\R(\x1,\x2) \,=} \\ &&
\left( {1\over 2} \ln(\x1) \ln(\x2)
                -\ln(\x1) \ln(1\!-\!\x1)
                +{1\over 2} \z2-\Li_{2}(\x1) \right) +
              (\x1\leftrightarrow\x2)\, .
\nn
\end{eqnarray}
In addition, it is convenient, to define the symmetric function
$\RR1(\x1,\x2)$, which contains a particular combination of 
multiple polylogarithms \cite{Goncharov}, 
\begin{eqnarray}
  \label{eq:RRfunctions}
\lefteqn{
\RR1(\x1,\x2) \,=\, \left(
          \ln(\x1) \Li_{1,1}\left({\x1 \over \x1\!+\!\x2},\x1\!+\!\x2 \right)
         -{1 \over 2} \z2 \ln(1\!-\!\x1\!-\!\x2)
\right. 
} \\ &&
\left.
         +\Li_{3}(\x1\!+\!\x2)
         -\ln(\x1) \Li_{2}(\x1\!+\!\x2)
         -{1 \over 2}\ln(\x1) \ln(\x2) \ln(1\!-\!\x1\!-\!\x2)
\right.
\nn \\ & &
\left.
         -\Li_{1,2}\left({\x1 \over \x1\!+\!\x2},\x1\!+\!\x2 \right)
         -\Li_{2,1}\left({\x1 \over \x1\!+\!\x2},\x1\!+\!\x2 \right)
       \right) + (\x1\leftrightarrow\x2)\, .
\nn
\end{eqnarray}

We have made the following checks on our result. 
As remarked, the infrared poles agree with the 
structure predicted by Catani \cite{Catani:1998bh}. 
This provides a strong check of the complete pole structure of our result.
In addition, we have tested various relations between the $c_i$. For instance,  
the combination $\x1  c_6$ is symmetric under 
exchange of $\x1$ with $\x2$.
Finally, we could compare with the result for the squared 
matrix elements, 
i.e. the interference of the two-loop amplitude with the Born amplitude, 
and the interference of the one-loop amplitude with itself. 
The results of \cite{Garland:2001tf} are given in terms of one- and
two-dimensional harmonic polylogarithms, which form a subset of the 
multiple polylogarithms \cite{Goncharov}.
Thus, we have performed the comparision analytically and we agree with 
the results of \cite{Garland:2001tf}. 

\section{Conclusions}
Our result represents one contribution to the 
full next-to-next-to-leading order calculation of 
$ e^+ e^- \rightarrow \mbox{3-jets}$. 
It has been obtained by means of an efficient method based on 
nested sums and is expressed in terms of multiple polylogarithms 
with simple arguments.
As a consequence, our result can be continued analytically and 
applies also to $(2+1)$-jet production in deep-inelastic scattering 
or to the production of a massive vector boson in hadron-hadron collisions.
At the same time, it provides an important cross check on the results for the
squared matrix elements \cite{Garland:2001tf} with a completely 
independent method.

After the results of section \ref{sec:calculation} had been presented at this 
conference, Garland et al. published results for the complete two-loop 
amplitude for $e^+e^-\to q \qb g$. Our results are in agreement with 
ref.~\cite{Garland:2002ak}.

\end{document}